\documentclass[epj]{svjour}
\usepackage{graphics,epsfig}

\newcommand{\bs}{\mathbf{s}}
\newcommand{\bb}{\mathbf{b}}
\newcommand{\half}{{\textstyle{1\over 2}}}

\begin{document}
\title{Chemical freeze-out temperature in hydrodynamical description 
       of Au+Au collisions at $\sqrt{s_{\mathrm{NN}}}=200$ GeV}
\titlerunning{Chemical freeze-out temperature in hydrodynamical description
              of heavy-ion collisions}
\author{Pasi Huovinen
\thanks{\emph{Present address:} Department of Physics, Purdue University,
                                West Lafayette, IN 47907, USA}
}                     
\institute{Department of Physics, University of Virginia, P.O.\ Box 400714,
           Charlottesville, VA 22904-4714, USA\\ \email{ph4h@virginia.edu}}
\date{\today}
%
\abstract{
We study the effect of separate chemical and kinetic freeze-outs to
the ideal hydrodynamical flow in Au+Au collisions at RHIC
($\sqrt{s_{\mathrm{NN}}}=200$ GeV energy). Unlike earlier studies we
explore how these effects can be counteracted by changes in the
initial state of the hydrodynamical evolution. We conclude that the
reproduction of pion, proton and antiproton yields necessitates a
chemical freeze-out temperature of $T\approx 150$ MeV instead of $T =
160$--170 MeV motivated by thermal models. Contrary to previous
reports, this lower temperature makes it possible to reproduce the
$p_T$-spectra of hadrons if one assumes very small initial time,
$\tau_0 = 0.2$ fm/$c$. However, the $p_T$-differential elliptic flow,
$v_2(p_T)$ remains badly reproduced. This points to the need to
include dissipative effects (viscosity) or some other refinement to
the model.
}

\maketitle

\section{Introduction}
   \label{intro}

Ideal fluid hydrodynamical models have been very successful in
describing the bulk behaviour of particles formed in heavy-ion
collisions at RHIC. The low-$p_T$ single particle spectra as well as
the transverse momentum dependence of elliptic anisotropy ($v_2(p_T)$)
are reproduced nicely~\cite{Huovinen:2006jp,Kolb:2003dz}. This success
has been one of the reasons to conclude that partonic state of matter
with exceptionally low shear viscosity has been formed at
RHIC~\cite{Gyulassy:2004zy}.

However, these results have been achieved using ideal fluid
hydrodynamical models, which assumed chemical equilibrium until the
very end of the evolution of the system. This assumption is
questionable, since the cross-sections of inelastic, particle number
changing processes, are smaller than the cross-sections of elastic and
quasi-elastic processes. Thus it is natural to assume that the system
would still maintain local kinetic equilibrium when it begins to
deviate from local chemical equilibrium.

This kind of approach is also supported by experimental data. The
final hadron abundances in Au+Au collisions at RHIC can be well
described by a hadron gas in approximate chemical equilibrium at
$T_\mathrm{ch}\approx 160$--175
MeV~\cite{BraunMunzinger:2003zd,Adams:2005dq}.  The reproduction of
the slopes of particle distributions using a blast-wave model requires
much lower temperatures around $T_\mathrm{kin}\approx 90$--130 MeV
(depending on centrality)~\cite{Adams:2003xp}. The hydrodynamical
models assuming chemical equilibrium usually require freeze-out
temperatures close to the blast-wave fits\footnote{With the notable
exception of refs.~\cite{Eskola:2002wx,Eskola:2005ue}.} and cannot
reproduce all observed particle yields simultaneously. It is thus
reasonable to postulate two separate freeze-outs: first a chemical
freeze-out where the yields of various particle species are fixed
(frozen out) and somewhat later a kinetic freeze-out where the
particles scatter for the last time and the momentum distributions
cease to evolve.

The formalism to describe such a chemically frozen hadron gas has been
known for quite a long time~\cite{Bebie:1991ij}. There have been
several applications of this formalism to the hydrodynamical
description of heavy ion collisions at SPS ($\sqrt{s_\mathrm{AA}}=17$
GeV) and RHIC ($\sqrt{s_\mathrm{AA}}=130$ and 200 GeV)
energies~\cite{Arbex:2001vx,Teaney:2002aj,Hirano:2002ds,Kolb:2002ve},
but the results have been unsatisfactory. The conclusion of these
studies has been that if the chemical composition of the hadron gas
freezes out at hadronization, an ideal fluid hydrodynamical model can
reproduce neither the single particle spectra nor the
$p_T$-differential anisotropy~\cite{Adcox:2004mh,Hirano:2005wx}.
However, these studies are lacking in such a sense that they used the
same initial state for both the chemically equilibrated and chemically
frozen description. It is known that $p_T$-distributions are sensitive
not only to the equation of state and kinetic freeze-out temperature,
but also to the initial pressure gradients, \textit{i.e.} to the
initial density distribution~\cite{Dumitru:1998es}. Here we redo the
hydrodynamic calculations once more to explore whether there is such
an initial state which leads to an acceptable reproduction of the
$p_T$-spectra and elliptic flow of pions and antiprotons.

\section{Equation of State}
  \label{EoS}

As a baseline, we use an equation of state (EoS) in chemical
equilibrium. We construct it in the usual way: The hadronic phase is
described by ideal resonance gas consisting of all hadrons and
resonances listed in the Particle Data Book~\cite{Eidelman:2004wy}
with mass below 2 GeV. The plasma phase is described by an ideal
massless parton gas consisting of gluons and three quark
flavours. There is a first order phase transition between these two
phases at $T_c = 170$ MeV. Note that this EoS is slightly different
from the EoS with a first order phase transition used in
ref.~\cite{Huovinen:2005gy} and EoS Q used in ref.~\cite{Kolb:2000sd}.
The phase transition temperature in this work is $T_c=170$ MeV instead
of 165 MeV and the number of quark flavours in the plasma phase is 3
instead of 2.5. The latter results in slightly larger latent heat. As
can be seen by comparing the results here and in
ref.~\cite{Huovinen:2005gy}, these differences have only a small
effect on the particle spectra and anisotropies. In the following we
refer to this EoS and the corresponding initial state as CE.

We construct an EoS out of chemical equilibrium in the way outlined in
ref.~\cite{Bebie:1991ij} and later applied in
refs.~\cite{Teaney:2002aj,Hirano:2002ds,Kolb:2002ve}. Below the
chemical freeze-out temperature all inelastic, particle number
changing processes have ceased, but elastic and quasielastic
scatterings are still frequent enough to keep the system in kinetic
equilibrium. The quasielastic scatterings lead to frequent formation
and decay of resonances, which means that the yields of resonances and
their daughter particles, say $\rho$-mesons and pions, still are in
relative chemical equilibrium. This approach is therefore called
partial chemical equilibrium (PCE). The presence of resonances means
that the actual particle number of any particle species is not
conserved after chemical freeze-out, but the effective particle number
is. The effective particle number is defined as $\bar N_i = N_i +
\sum_j n_j^{(i)}N_j$, where $N_i$ is the actual number of particle
species $i$, $n_j^{(i)}$ is the number of particles $i$ formed in the
decay of resonance $j$ including the branching ratios, and $N_j$ is
the number of resonances $j$. The sum is over all the resonances with
lifetimes smaller than the characteristic lifetime of the system. We
take the characteristic lifetime to be 10 fm/$c$ irrespective of the
actual lifetime of the system. The following particles and resonances
have lifetimes longer than 10 fm/$c$ and are thus considered stable:

\[ \pi,\,K,\,\eta,\,\omega,\,p,\,n,\,\eta',\,\phi,\,\Lambda,\,\Sigma,\,
   \Xi,\,\Lambda(1520),\,\Xi(1530),\,\Omega,
\]
and the corresponding antiparticles. The corresponding effective
potential of resonances is obtained in a similar fashion:
$\mu_j = \sum_i n_j^{(i)} \mu_i$, where the sum is over all decay
products of the resonance $j$.

\begin{figure}
  \begin{center}
   \epsfxsize 6cm \epsfbox{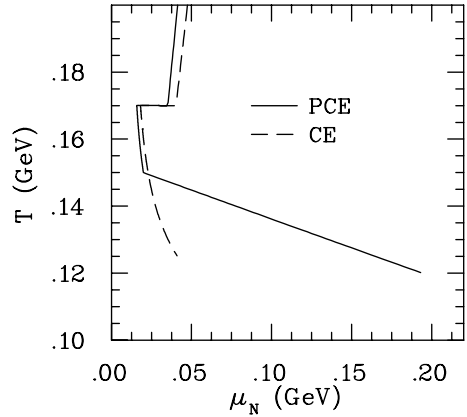}
  \end{center}
 \caption{Isentropic thermodynamic trajectories below 200 MeV at
          RHIC. PCE corresponds to chemically frozen calculation and
          CE corresponds to the center of the system in chemically
          equilibrated calculation. Above the phase transition
          temperature nucleon chemical potential $\mu_N = 3\mu_q$. The
          curves terminate at the kinetic freeze-out temperature
          ($\langle T_{\mathrm{kin}}\rangle=120$ and 125 MeV). Note
          that the difference above the chemical freeze-out temperature
          $T_\mathrm{ch} = 150$ MeV is due to different initial
          states, see sect.~\ref{init}. For PCE and CE, $s/n_B=453.5$
          and 395, respectively, where $n_B$ is net baryon density.}
 \label{T-mu-plane}
\end{figure}

We treat each isospin state of each particle species independently.
This results in 35 conserved quantities and chemical potentials in the
chemically frozen stage. Tabulating the EoS as a function of 36
quantities (35 chemical potentials and temperature) is unpractical,
but avoidable by taking the advantage of the isentropic nature of
ideal fluid hydrodynamics. Since entropy is conserved, the ratio of
the effective particle number density ($\bar{n}_i =\bar{N}_i/V$) and
entropy density, $\bar{n}_i/s$, stays constant on streamlines of the
flow. The expansion thus traces a trajectory of constant $\bar{n}_i/s$
in $(T,\{\mu_i\})$ space and we need to evaluate the EoS only on this
trajectory. The trajectories in the $(T,\mu_N)$ plane of the
chemically frozen (PCE) and equilibrated systems (CE) at RHIC are
shown in Fig.~\ref{T-mu-plane}.

Qualitatively our EoS is similar to the EoSs in
refs.~\cite{Teaney:2002aj,Hirano:2002ds,Kolb:2002ve}. The relation
between pressure and energy density is almost identical in chemically
equilibrated and frozen cases whereas at fixed energy density the
chemically frozen system is much colder than a system in chemical
equilibrium.

\section{Initial conditions}
  \label{init}

\begin{table}[b]
\caption{Initial time, phase transition temperature, chemical
         freeze-out temperature, kinetic freeze-out energy density and
         the corresponding average temperature on kinetic freeze-out
         surface used in chemical equilibrium (CE) and partial chemical
	 equilibrium (PCE) calculations.}
\label{paratable}       
\begin{tabular}{lll}
\hline\noalign{\smallskip}
                           & CE  & PCE  \\
\noalign{\smallskip}\hline\noalign{\smallskip}
$\tau_0$ (fm/$c$)          & 0.6 & 0.2 \\
$T_{\mathrm{c}}$ (MeV)     & 170 & 170 \\
$T_{\mathrm{ch}}$ (MeV)    & --  & 150 \\
$\epsilon_{\mathrm{kin}}$ (GeV/fm$^3$)  & 0.065 & 0.117 \\
$\langle T_{\mathrm{kin}}\rangle$ (MeV) & 125 & 120 \\
\noalign{\smallskip}\hline
\end{tabular}
\end{table}

\begin{table*}
\caption{Measured and calculated particle yields at midrapidity in
         most central (0--5\%) Au+Au collisions at
         $\sqrt{s_{\mathrm{NN}}}=200$ GeV. CE stands for chemical
         equilibrium, $T_{\mathrm{ch}}=170$ and 150 MeV denote cases
         where chemical freeze-out takes places at 170 and 150 MeV
         temperatures, respectively, but the initial state of the
         system is similar to the case CE. PCE is the final partial
         chemical equilibrium result where chemical freeze-out takes
         place at $T_{\mathrm{ch}}=150$ MeV and the initial conditions
         are adjusted to reproduce the observed $p_T$-distributions.}
\label{yieldtable}       
\begin{tabular}{lcccccc}
\hline\noalign{\smallskip}
 $\mathrm{d}N/\mathrm{d}y$
                      & $\pi^+$ & $\pi^-$ & $K^+$ & $K^-$ & $p$ & $\bar{p}$\\
\noalign{\smallskip}\hline\noalign{\smallskip}
PHENIX~\cite{Adler:2003cb} 
                      & $286.4\pm 24.2$ 
                                & $281.8\pm 22.8$
                                          & $48.9\pm 5.2$
                                                  & $45.7\pm 5.2$
                                                          & $18.4\pm 2.6$
                                                             & $13.5\pm 1.8$ \\
CE                        & 273 & 273     & 44    & 42    & 10 &  5       \\
$T_{\mathrm{ch}}=170$ MeV & 268 & 268     & 52    & 50    & 25 & 20       \\
$T_{\mathrm{ch}}=150$ MeV & 265 & 265     & 51    & 49    & 18 & 13       \\
PCE                       & 278 & 278     & 53    & 51    & 18 & 14       \\
\noalign{\smallskip}\hline
\end{tabular}
\end{table*}

We use the same boost-invariant hydrodynamic code as in
ref.~\cite{Huovinen:2005gy}. In the case of chemical equilibrium (CE)
we use the same initial conditions as in ref.~\cite{Huovinen:2005gy}
for the equation of state with a first order phase transition: The
initial entropy density distribution is a linear combination of the
density of participants and binary collisions in the transverse plane
whereas the initial baryon density is proportional to the number of
participants. The parameter values are fixed to reproduce the
$p_T$-spectra of pions and net protons ($p - \bar{p}$) in the most
central collisions and the centrality dependence of charged hadron
multiplicity at midrapidity. The initial time, phase transition
temperature and freeze-out criteria for cases CE and PCE are shown in
table~\ref{paratable}. Note that the freeze-out energy density is
slightly lower than in the corresponding calculation of
ref.~\cite{Huovinen:2005gy} because the EoS is slightly different (see
sect.~\ref{EoS}).

We find that the same initialization will not work in the chemically
frozen case. To reproduce the slopes of the $p_T$-distributions we
increase transverse flow by making the initial pressure gradients
steeper and by starting the hydrodynamical evolution earlier. An
acceptable result is achieved when the initial entropy density
distribution is assumed to be proportional to the number of binary
collisions in the transverse plane and the initial time is taken to be
$\tau_0 = 0.2$ fm/$c$.

This choice of initial time is bold since it requires hydrodynamics to
be applicable almost immediately after the formation time of the
partons of the system, but is nevertheless
plausible~\cite{Huovinen:2006jp,Eskola:2005ue}. First, in the
pQCD+saturation method for calculating the particle production of
ref.~\cite{Eskola:2005ue}, the average energy and particle number
densities are almost equal to those of a system of fixed temperature.
Thus there is no need for particle number changing processes to
achieve thermalization. Second, for massless particles, the relation
between pressure and energy density is $\epsilon = 3 P$ for any
isotropic momentum distribution. Thus the use of hydrodynamics to
describe the build-up of collective flow could be a reasonable
assumption even if the momentum distribution differs from thermal
equilibrium distribution. Third, this ideal gas EoS may be applicable
very rapidly, since isotropization of the momentum distribution occurs
much faster than thermal
equilibration~\cite{Berges:2005ai,Kovchegov:2007pq}. Finally, the
studies of plasma instabilities support the notion of fast
thermalization of the system~\cite{Mrowczynski:2005ki}.

If the proportionality between the initial entropy density and the
number of binary collisions is independent of the centrality of the
collision, the centrality dependence of the final particle
multiplicity is not reproduced~\cite{Kolb:2001qz}. To correct this, we
modify the parametrization given in ref.~\cite{Kolb:2001qz} by
assuming an impact parameter dependent proportionality constant:
\begin{eqnarray}
  K_s(\tau_0=0.2\,\mathrm{fm};b)  
   & = & 0.26942\,\mathrm{fm}^{-6}\,b^3 + 10.9\,\mathrm{fm}^{-4}\,b \nonumber\\
   & + & 453.5\,\mathrm{fm}^{-3}, \nonumber
\end{eqnarray}
and parametrize the initial entropy density in the transverse plane as
 \[
   s(\bs;\tau_0;\bb) = K_s(\tau_0;b) \,
   T_A\left(\bs{+}\half\bb\right) T_B\left(\bs{-}\half\bb\right)\,,
 \]
where $T_A$ is the customary nuclear thickness
function~\cite{Kolb:2001qz}. The initial baryon number distribution is
obtained in the same way, \textit{i.e.} it is taken to be proportional
to the initial entropy density.

\section{Particle yields and chemical freeze-out temperature}
 \label{cfoT}

When separate chemical and kinetic freeze-outs are discussed in recent
literature, chemical freeze-out is assumed to take place immediately
after hadronization at $T_\mathrm{ch} = 160$--170
MeV~\cite{Arbex:2001vx,Teaney:2002aj,Hirano:2002ds,Kolb:2002ve}. Either
the ratios of all particle spe\-cies are taken to be fixed at this
temperature~\cite{Teaney:2002aj,Hirano:2002ds,Kolb:2002ve} or only
strange particle yields are supposed to
freeze-out~\cite{Arbex:2001vx}.  The idea of chemical freeze-out at
hadronization is conceptually
attractive~\cite{Muller:2005wi,Heinz:2006ur} and to some extent
supported by thermal models which lead to
temperatures~\cite{BraunMunzinger:2003zd,Adams:2005dq} quite similar
to the predicted phase transition temperature~\cite{Karsch}. However,
in the context of hydrodynamical model, we have found that assuming
chemical freeze-out of all particle species at $T\approx 170$ MeV,
leads to proton and antiproton yields that are too large when the
model is tuned to reproduce the observed pion multiplicity. In most
central collisions we obtain $\mathrm{d}N_p/\mathrm{d}y = 25$ protons
and $\mathrm{d}N_{\bar{p}}/\mathrm{d}y=20$ antiprotons at midrapidity
instead of the experimentally observed yields of
$\mathrm{d}N_p/\mathrm{d}y = 18.4 \pm 2.6$ and
$\mathrm{d}N_{\bar{p}}/\mathrm{d}y = 13.5 \pm 1.8$ (see
Table~\ref{yieldtable} and fig.~\ref{ptcentral}, case 
$T_\mathrm{ch} = 170$ MeV). This discrepancy is not surprising. Some
thermal model fits to RHIC data have actually led to temperatures near
or below 160 MeV rather than 170
MeV~\cite{Kaneta:2004zr,Andronic:2005yp}. Thermal models also tend to
lead to $p/\pi$ ratios which are slightly larger than the experimental
ratio, although still within experimental
errors~\cite{Kaneta:2004zr,Andronic:2005yp,Baran:2003nm,Cleymans:2004pp}.
If the model is required to fit only $K/\pi$ and $p/\pi$ ratios, the
temperature is lower, $T \approx 152$ MeV~\cite{Cleymans:2004pp}.

As was argued in~\cite{Hirano:2005wx}, the evolution of the average
transverse momentum per particle, and thus the observed $p_T$ spectra,
is sensitive to pion number changing processes. In a hydrodynamical
model it is thus important to describe the pion chemistry correctly
and get the pion chemical freeze-out temperature right even if that
means a worse reproduction of strange baryon yields. Inspired by the
results of ref.~\cite{Kaneta:2004zr}, we take $T=150$ MeV as the
temperature where the pion number changing processes become negligible
and the effective pion number freezes out. We expect that the better
reproduction of the pion to proton and the pion to antiproton ratios
gives a more realistic description of the temperature evolution and
therefore of the evolution of particle distributions.
As can be seen in the case $T_\mathrm{ch}=150$ MeV of
Table~\ref{yieldtable}, this approach gives the expected results:
pion, kaon, proton and antiproton yields are almost exactly
reproduced. The strange baryon yields, on the other hand, become too
small at $T_{\mathrm{ch}}=150$ MeV. 

The possible contribution from weak decays of strange particles causes
a significant uncertainty in determining the chemical freeze-out
temperature~\cite{BraunMunzinger:2003zd}. According to the PHENIX
collaboration~\cite{Adler:2003cb,Roy}, the data has been corrected for
the feed-down from weak decays of $\Lambda$'s and $\Sigma^0$'s, but
whether the pion spectrum contains a contribution from $K^0_S$ decays,
is not clearly explained. In these calculations we have assumed that
there is no contribution from any weak decays in the PHENIX data. The
decay of thermally produced $K^0_S$'s and $\Sigma^\pm$'s would
increase the pion yield at midrapidity by $\sim 40$ pions and one
proton/antiproton. For the cases $T_{\mathrm{ch}}=170$ and 150 MeV,
the pion yield is thus at the experimental upper limit. The reduction
of initial entropy by 5-10\% would bring the pion yield closer to the
experimental value and reduce the proton/antiproton yield by the same
5-10\%. The proton and antiproton yields at $T_{\mathrm{ch}}=170$ MeV
temperature would be too large in this case too. At
$T_{\mathrm{ch}}=150$ MeV, the proton and antiproton yields would be
within experimental errors, but one can argue that slightly larger
freeze-out temperature $T_{\mathrm{ch}}\approx 155$ MeV is favoured by
the data.

In the case PCE the same procedure leads to pion yield larger than the
data, which necessitates a larger adjustment in the initial
entropy. However, after the adjustment, the conclusion is the same
than for the cases $T_{\mathrm{ch}}=170$ and 150 MeV: Even if the pion
yield contains a contribution from $K^0_S$ decays, $T_{\mathrm{ch}}=170$
MeV chemical freeze-out temperature leads to too large proton and
antiproton yields and a temperature around $T_{\mathrm{ch}}=150$--155
MeV is favoured.

On the other hand, if the experimental correction for weak decays is
not perfect and the spectra contains feed-down from strange baryons,
the relative increase in the proton/antiproton yields would be larger
than in the pion yield. In such a case the favoured freeze-out
temperature would be even smaller than what is suggested here.

\section{Transverse momentum spectra}
  \label{pT}

\begin{figure}
  \begin{center}
   \epsfxsize 6cm \epsfbox{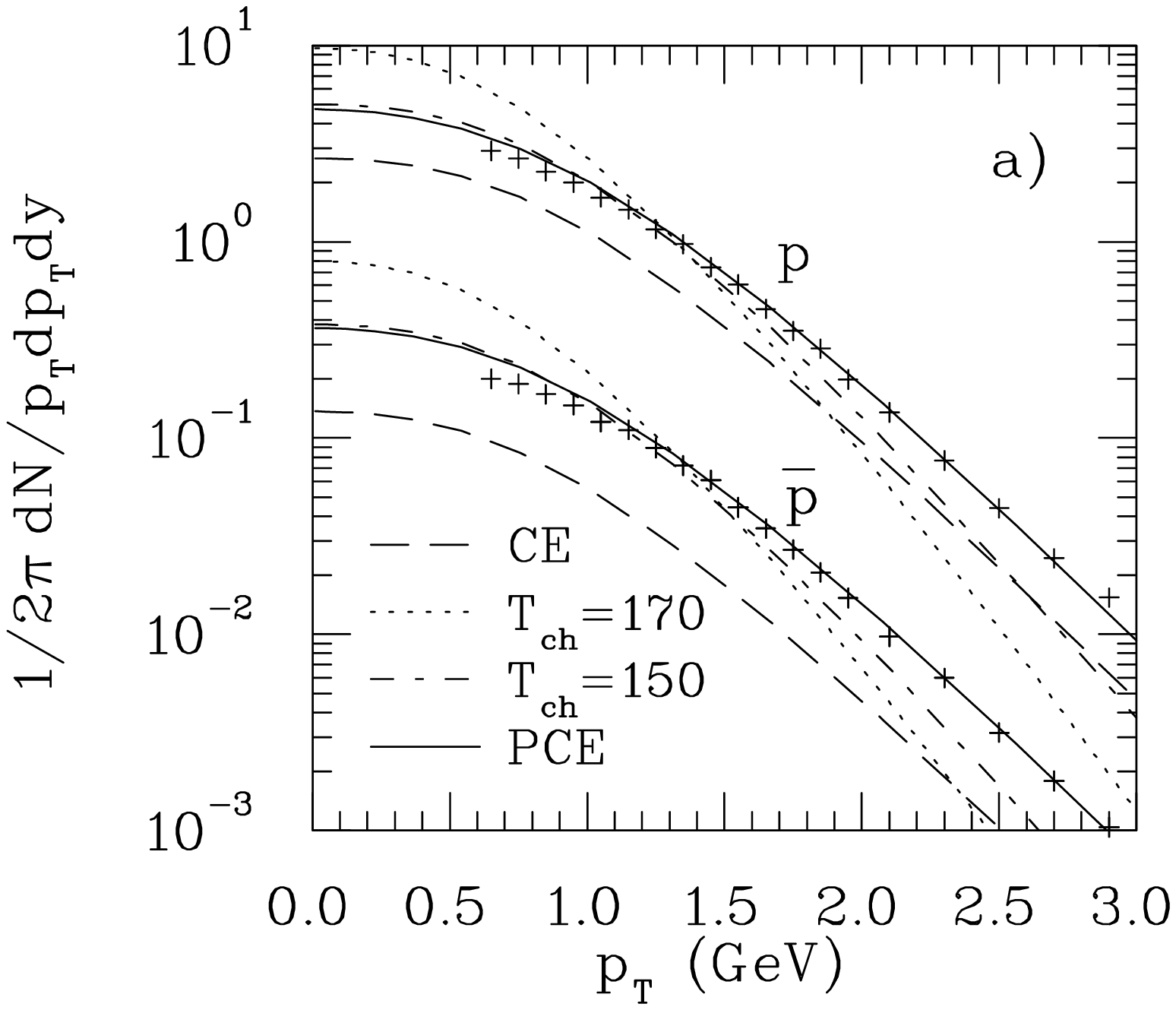}
   \epsfxsize 6cm \epsfbox{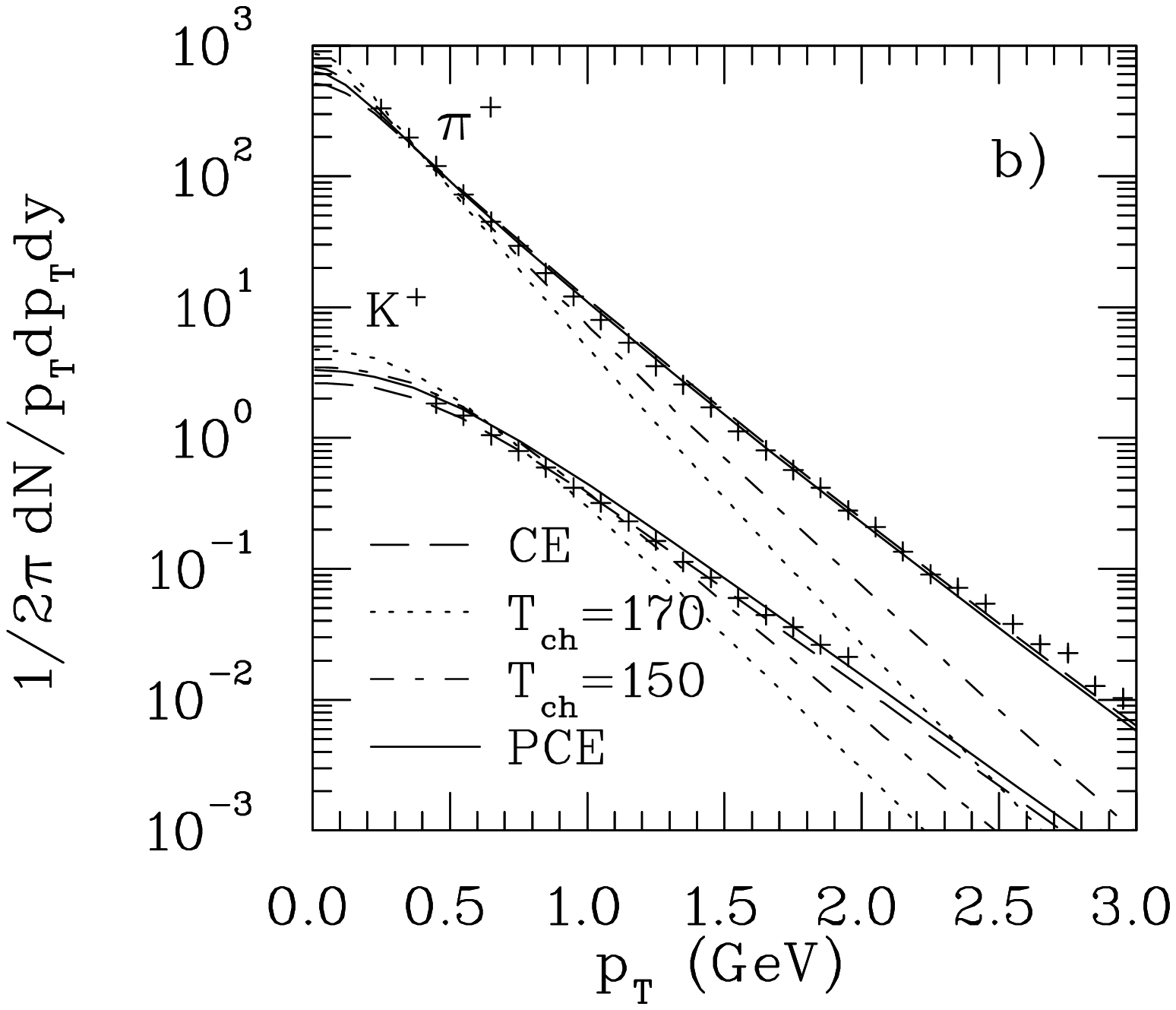}
  \end{center}
 \caption{(a) Proton and antiproton and (b) pion ($\pi^+$) and kaon
          ($K^+$) $p_T$-spectra in most central (0--5\%) Au+Au
          collisions at $\sqrt{s_{\mathrm{NN}}}=200$ GeV compared with
          hydrodynamical calculations using different chemical
          freeze-out descriptions (see the text). The data was taken
          by the PHENIX collaboration~\cite{Adler:2003cb}. For clarity
          the antiproton and kaon spectra are scaled by a factor
          $10^{-1}$.}
 \label{ptcentral}
\end{figure}

The transverse momentum distributions for protons, antiprotons, pions
and kaons in the most central collisions are shown in
fig.~\ref{ptcentral}. As before~\cite{Huovinen:2005gy}, the
calculation with chemical equilibrium (CE) reproduces the pion and
kaon data well, but it reproduces only the slopes of proton and
antiproton distributions. If we proceed as in
refs.~\cite{Hirano:2002ds,Kolb:2002ve} and replace the EoS by an EoS
with partial chemical equilibrium below $T_\mathrm{ch}=170$ MeV,
adjust only the kinetic freeze-out density,\footnote{Note that in the
actual calculation we use constant energy density as freeze-out
criterion for numerical simplicity. In the following we talk about
freeze-out temperature instead.} and keep everything else unchanged,
the result is a disaster ($T_\mathrm{ch}=170$ in the figure). Even at
$\langle T_\mathrm{kin}\rangle=100$ MeV average freeze-out
temperature, the slopes of proton and antiproton spectra are far too
steep. Decreasing the freeze-out temperature/density does not help the
overall fit, since the pion spectrum becomes steeper with decreasing
freeze-out temperature. This slightly counterintuitive behaviour was
explained in ref.~\cite{Hirano:2005wx}. In boost-invariant expansion,
transverse energy per unit rapidity, $\mathrm{d}E_T/\mathrm{d}y$,
decreases with increasing time. In partial chemical equilibrium
particle number is conserved and transverse energy has to be
distributed among the same number of particles. Consequently $\langle
p_T\rangle$ decreases and the slope of the $p_T$ distribution
steepens.

Another problem with $T_\mathrm{ch}=170$ MeV temperature can also be
seen in fig.~\ref{ptcentral}: There are too many protons and
antiprotons as discussed in the previous section. This can be cured by
decreasing the chemical freeze-out temperature to $T_\mathrm{ch}=150$
MeV. The calculation labelled $T_\mathrm{ch}=150$ in
fig.~\ref{ptcentral} is performed in the same way as
$T_\mathrm{ch}=170$.  Only the EoS is changed, but the initial state
is kept the same. When the kinetic freeze-out temperature is kept the
same, $\langle T_\mathrm{kin}\rangle=100$ MeV, as for
$T_\mathrm{ch}=170$, the particle distributions are closer to the
data, but the slopes are still too steep. The flattening of the
spectra is easy to understand.  Between $T=170$ and 150 MeV particle
number is not conserved but the energy stored in masses of heavy
particles is converted to kinetic energy of light particles (mostly
pions) when the system cools. Thus $\langle p_T\rangle$ may increase
even if $\mathrm{d}E_T/\mathrm{d}y$ decreases, and $\langle
p_T\rangle$ is larger at the time of chemical freeze-out when the
particle number is fixed and pion $\langle p_T\rangle$ begins its slow
decrease.

\begin{figure}
  \begin{center}
   \epsfxsize 6cm \epsfbox{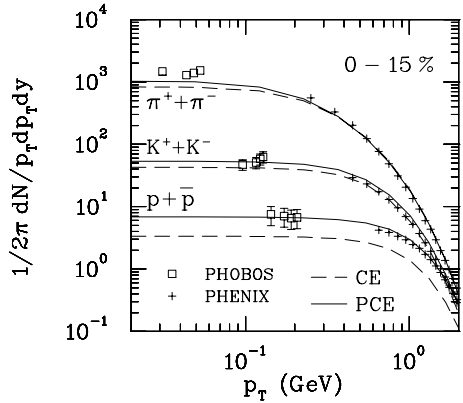}
  \end{center}
 \caption{Transverse momentum spectra of charged pions, charged kaons
          and protons + antiprotons in 15\% most central Au+Au
          collisions at $\sqrt{s_{\mathrm{NN}}}=200$ GeV in the
          ultralow-$p_T$ region measured by PHOBOS~\cite{Back:2004zx}
          compared with hydrodynamical calculations. The systematic
          and statistical errors of the PHOBOS data have been added in
          quadrature. The PHENIX data~\cite{Adler:2003cb} are an
          average of the data in 0-5\%, 5-10\% and 10-15\% centrality
          bins and have only statistical errors.}
 \label{low}
\end{figure}

Neither the steeper initial profile given by the pure binary collision
profile nor the small initial time $\tau_0 = 0.2$ fm/$c$ is sufficient
alone to create large enough flow before chemical freeze-out to fit
the data. When these two are used together (PCE), the fit to the data
is good. The fit to pion data is even better than in the chemical
equilibrium case (CE) at low $p_T$, where PCE depicts a concave
curvature typical for a finite pion chemical
potential~\cite{Kataja:1990tp}. At low $p_T$ the calculation still
suffers from excess of protons and especially antiprotons as if even a
larger flow velocity and a smaller chemical freeze-out temperature
were necessary. We have checked that the binary collision profile and
short initial time combined with an EoS with $T_\mathrm{ch} = 170$ MeV
does not lead to satisfactory reproduction of the slopes either, but
the extra push created in the hadronic equilibrium stage between
$T=170$ and 150 MeV temperatures is necessary to fit the data.

\begin{figure}
  \begin{center}
   \epsfxsize 6cm \epsfbox{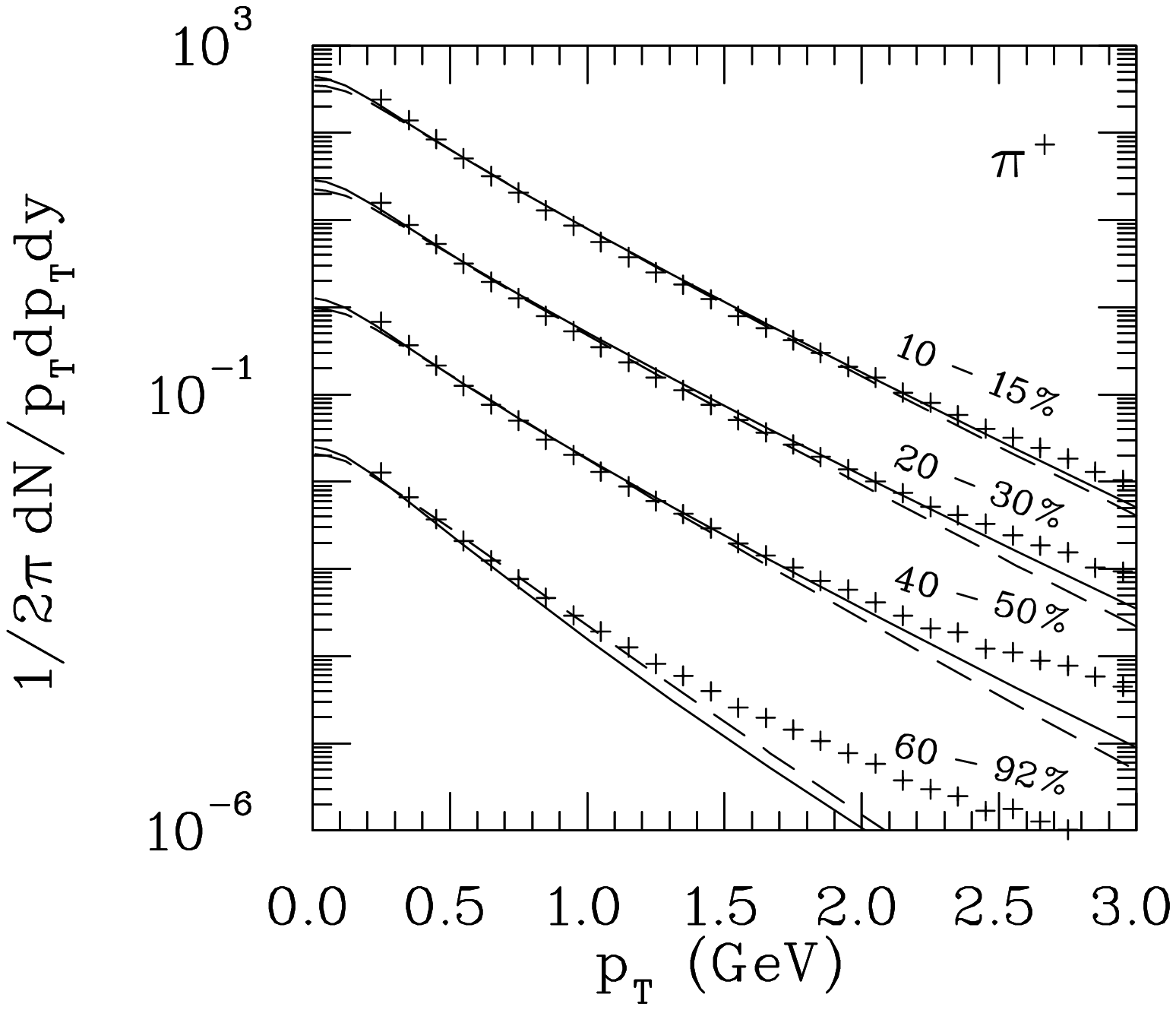}
   \epsfxsize 6cm \epsfbox{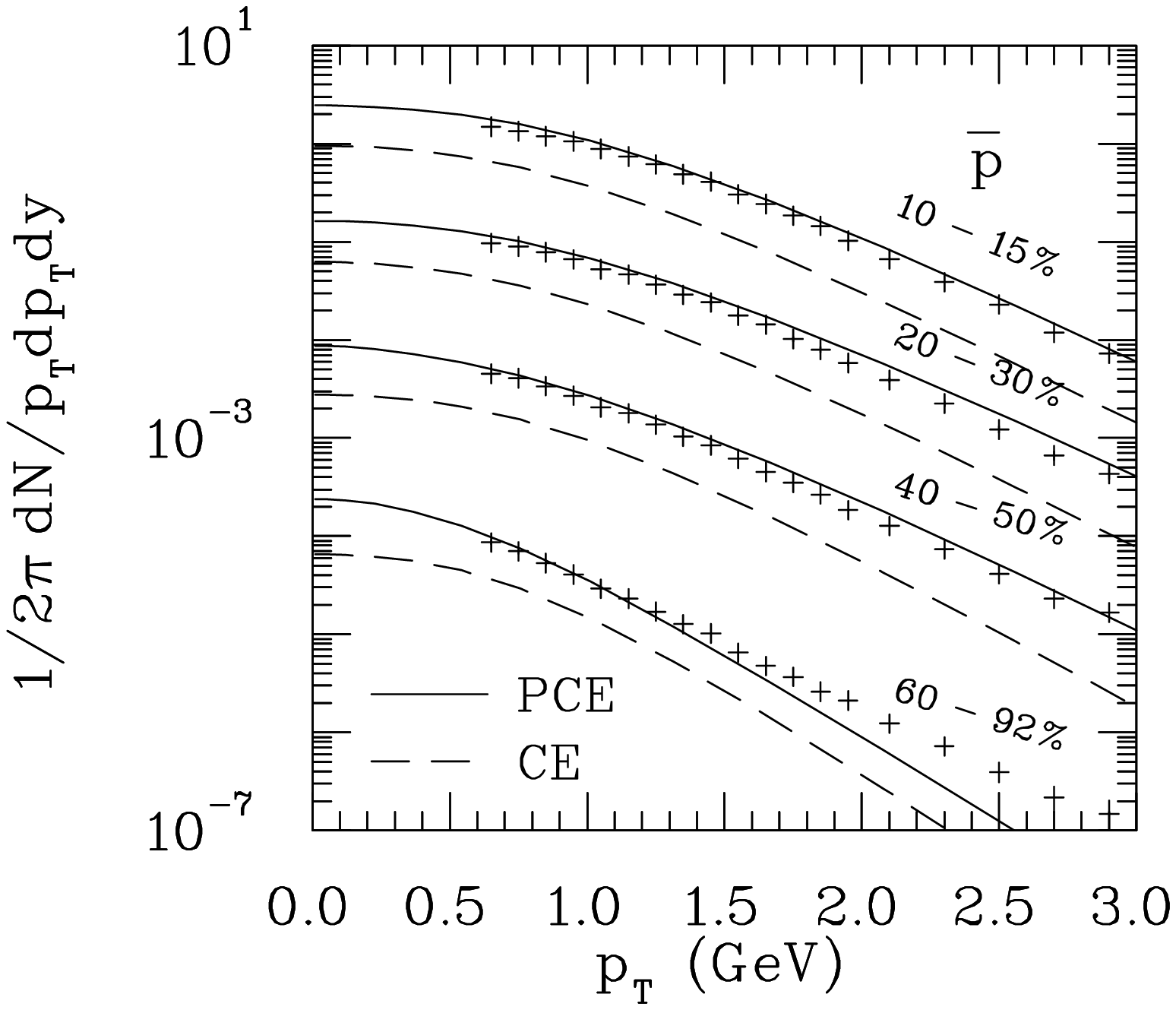}
  \end{center}
 \caption{Pion ($\pi^+$) and antiproton ($\bar{p}$) $p_T$-spectra in
          semi-central to peripheral Au+Au collisions at
          $\sqrt{s_{AA}} = 200$ GeV compared with hydrodynamical
          calculations (see text). The data was taken by the PHENIX
          collaboration~\cite{Adler:2003cb}. For clarity the spectra at
          centralities 20 - 30\%, 40 - 60\% and 60 - 80\% are scaled
          by factors 10$^{-1}$, 10$^{-2}$ and 10$^{-3}$,
          respectively.}
 \label{ptsemi}
\end{figure}

Figure~\ref{low} shows the results for the very low transverse
momentum measured by the PHOBOS collaboration~\cite{Back:2004zx} as
well as the spectra measured by the PHENIX
collaboration~\cite{Adler:2003cb} at larger $p_T$. The model PCE works
well also at this $p_T$ region reproducing the flat behaviour of the
data. The spectrum of pions is slightly below the data. This was seen
already in table~\ref{yieldtable} where pion multiplicity was shown to
be slightly below the data but still within the experimental
error. The increase in the initial entropy of the system to increase
pion multiplicity would, however, necessitate even lower chemical
freeze-out temperature not to exceed the observed kaon multiplicity.

Pion and antiproton $p_T$-spectra at various centralities are shown in
fig.~\ref{ptsemi}. The pattern is familiar from earlier hydrodynamical
studies: The more peripheral the collisions, the narrower the $p_T$
range where hydrodynamics can reproduce the spectra.  An interesting
detail is that antiproton spectra in semi-central collisions (20-30\%
and 40-50\% of total cross section) is slightly too flat. As mentioned
in the introduction, blast wave fits give higher kinetic freeze-out
temperatures in peripheral than in central collisions. Hydrodynamical
models have traditionally used fixed freeze-out temperature/density
for simplicity, but the overshooting of antiproton spectra here
suggests that a slightly larger kinetic freeze-out density in
semi-central than in central collisions might lead to better results.

\section{Elliptic anisotropy}
  \label{v2}

Unfortunately the success in the previous section cannot be repeated
for elliptic anisotropy. Figure~\ref{v2} shows the $p_T$-differential
elliptic anisotropy of pions and antiprotons in minimum bias Au+Au
collisions. The trend is similar to that in ref.~\cite{Hirano:2002ds}.
The chemical equilibrium result (CE) reproduces the pion data
excellently and is slightly above the proton data. Partial chemical
equilibrium (PCE) leads to a slope of the pion anisotropy that is too
large and the calculated result is clearly above the data. The proton
anisotropy is reproduced at $p_T < 600$ MeV, but increases too fast at
larger momenta.

\begin{figure}
  \begin{center}
   \epsfxsize 6cm \epsfbox{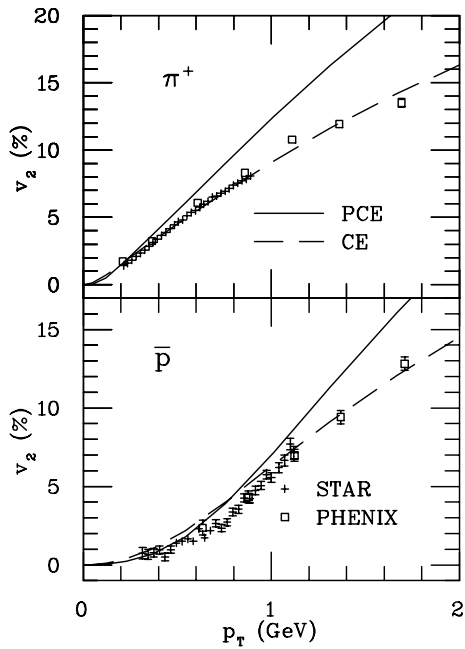}
  \end{center}
 \caption{Elliptic anisotropy of pions and antiprotons vs. transverse
          momentum in minimum bias Au+Au collisions at
          $\sqrt{s_{\mathrm{NN}}}=200$ GeV calculated using different
          chemical freeze-out descriptions (see text) and compared
          with the data by the STAR~\cite{Adams:2004bi} and
          PHENIX~\cite{Phenix-v2} collaborations.}
 \label{v2figure}
\end{figure}

It is worth noticing that the different $p_T$-differential $v_2$ is
mostly due to the different description of the hadron gas. As shown in
ref.~\cite{Hirano:2002ds}, the same initial state leads to different
$v_2$ in chemically equilibrated and frozen cases. The basic reason
for this is in the different temperature evolution in the hadron
gas. When both systems have reached the same density, the collective
flow field and its anisotropies are almost identical in both
cases. However, the chemically equilibrated system is hotter and thus
the random thermal motion is stronger. Thermal motion smears away the
underlying anisotropy of the collective flow field, and, therefore,
the final $v_2$ of particles of the chemically equilibrated system is
smaller than the $v_2$ of particles of the corresponding chemically
frozen system\footnote{For a more detailed discussion see
                       ref.~\cite{Hirano:2005wx}.}.

Comparison of these results with the recent results obtained using
viscous
hydrodynamics~\cite{Chaudhuri:2007zm,Romatschke:2007mq,Song:2007fn} is
particularly interesting. Dissipation reduces elliptic flow and it was
claimed in ref.~\cite{Song:2007fn} that even the postulated minimal
shear viscosity $\frac{\eta}{s} = \frac{1}{4\pi}$ causes very large
suppression. As shown here, suppression of elliptic flow is required
to fit the data when chemical equilibrium is lost in the hadron
gas. However, the actual size of the allowed viscous correction is
unknown since the EoS affects especially the proton anisotropy.  At
least in the chemically equilibrated case, the lattice QCD based EoS
leads to even larger proton $v_2(p_T)$ than the EoS used
here~\cite{Huovinen:2005gy}. Whether finite shear viscosity leads to a
correct reproduction of the data remains to be seen as well as how
large of a viscosity is allowed in plasma and hadronic
phases~\cite{Hirano:2005xf}.

\section{Summary and discussion}
  \label{summa}

Contrary to previous
reports~\cite{Hirano:2002ds,Kolb:2002ve,Adcox:2004mh,Hirano:2005wx},
we have shown that it is possible to reproduce both (nonstrange)
hadron yields and their transverse momentum spectra using ideal fluid
hydrodynamics with separate chemical and kinetic freeze-outs. The
difference between our approach and the previous studies is that we
used a different initial state for the chemically frozen system than
for the system in chemical equilibrium, adjusted the initial time to
be as small as $\tau_0=0.2$ fm/$c$ and changed the chemical freeze-out
temperature to be below the phase transition temperature to
$T_\mathrm{ch}=150$ MeV. However, our success is incomplete. Even
after these adjustments of the model, we were not able to reproduce
the $p_T$-differential elliptic flow of pions and protons. To describe
particle yields, their $p_T$-distributions and anisotropies, the
description of expansion has to be amended, most probably by including
dissipative effects: viscosity and diffusion.

Nevertheless, our results demonstrate how sensitive the expansion
dynamics during the hadronic phase is to hadron chemistry. Good
knowledge of hadron chemistry is therefore essential if we wish to
extract the viscosity of the QGP from the experimental data using
viscous hydrodynamics. The recent method of choice for this purpose
has been the so called hybrid hydro+cascade model where the expansion
is described using hydrodynamics until the system is hadronized and
the hadronic phase is described using hadronic
cascade~\cite{Hirano:2005xf,Bass:2000ib,Teaney:2001av,Nonaka:2006yn}.
These models have the advantage that the dissipative effects and
hadronic chemistry are included in the model and the kinetic
freeze-out is a result of hadronic cross sections without any free
parameters. So far cascade models used in these hybrid models have
been limited to two particle scatterings, but it has been argued that
multiparticle processes are essential for reproducing the proton and
antiproton yields~\cite{Rapp:2000gy,Cassing:2001ds}. It is therefore
not surprising that hydro+cascade models may have difficulties in
reproducing antiproton spectrum at the lowest values of transverse
momentum~\cite{Nonaka:2006yn}. On the other hand, multiparticle
processes are included in hydrodynamical models and as argued in
sect.~\ref{pT}, processes where heavy particles annihilate and form
lighter particles, and vice versa, are important for the build-up of
flow. For this reason we consider that a hydrodynamical description of
the hadronic phase is still worth pursuing as a complementary tool to
hybrid models, even if there are transport models where the
multiparticle processes are included~\cite{Cassing:2001ds}.

Another conclusion of these results is that the final spectra are
sensitive to the details of the initial state. As already stated in
ref.~\cite{Hirano:2005xf}, theoretical constraints to the initial
state and especially to its shape are of utmost importance if we wish
to extract the properties of the QGP from experimental data.

Some arguments favouring the use of hydrodynamics already at initial
time $\tau_0 = 0.2$ fm/$c$ were given in sect.~\ref{init}. We admit
that in a certain sense our result is similar to
ref.~\cite{Kolb:2002ve}, where Kolb and Rapp claimed that it is not
possible to fit the $p_T$-spectra without pre-hydrodynamic transverse
flow. Some of our arguments favouring the use of hydrodynamics at
$\tau=0.2$ fm/$c$ centered on the fact that for the relation 
$\epsilon = 3P$ to hold, it is sufficient that the momentum
distribution is isotropic. It does not need to be thermal. Thus one
may say that we are using hydrodynamics to calculate the
pre-hydrodynamic transverse flow, although the flow field in our case
is different from the one used in ref.~\cite{Kolb:2002ve}.

The low chemical freeze-out temperature $T_{\mathrm{ch}}=150$ MeV
clearly below the phase transition temperature $T_c=170$ MeV leads to
another problem. The chemical freeze-out temperature is almost
independent of centrality~\cite{Cleymans:2004pp}. It was argued in
ref.~\cite{Heinz:2006ur} that such an independence requires either
chemical freeze-out at phase transition or extremely steep temperature
dependence of the inelastic scattering rate. One way to solve this
apparent discrepancy is to assume that the chemical equilibration in
the hadron gas close to $T_c$ is very fast indeed due to massive
Hagedorn resonances~\cite{NoronhaHostler:2007fg}. These resonances
couple to multi-pion states and baryon-antibaryon pairs and thus lead
to fast chemical equilibration, but they appear only close to the
critical temperature.

The treatment of hadronic chemistry and choice of chemical freeze-out
used in this paper is by no means the final one. Since the data seem
to favour the notion that strange baryon yields are fixed at higher
temperature than the yields of pions, kaons and nucleons/antinucleons,
it would be relatively easy to use two separate chemical freeze-out
temperatures: one for strange baryons and another for everything else.
Another relatively straightforward improvement is to redo the
calculation of the chemical relaxation rate of
pions~\cite{Pratt:1999ku} for the RHIC environment and to use it to
dynamically find the local chemical freeze-out temperature by
comparing this rate to the expansion rate. This approach is
particularly appealing if the kinetic freeze-out surface is also found
by comparing the scattering and expansion rates. Finally, it would be
useful to calculate the particle yields dynamically during the
evolution using actual rate equations for particle number changing
processes. For baryons, preliminary study was already done in
ref.~\cite{Huovinen:2003sa}, but in that work mesons were supposed to
be in chemical equilibrium and the EoS was independent of baryon and
antibaryon densities. A proper calculation would require dynamical
treatment of mesons and calculating the EoS using actual particle
densities.

\begin{acknowledgement}
I thank K.~Haglin, T.~Hirano, H.~Honkanen, S.~Pratt, P.V.~Ruus\-kanen,
S.S.~R\"as\"anen and G.~Torrieri for many enlightening
discussions. Financial support from Johannes R\aa ttendahl foundation
is also gratefully acknowledged.
\end{acknowledgement}


\begin{thebibliography}{99}

\bibitem{Huovinen:2006jp}
  P.~Huovinen and P.~V.~Ruuskanen,
  Ann.\ Rev.\ Nucl.\ Part.\ Sci.\  {\bf 56} (2006) 163
  [arXiv:nucl-th/0605008].

\bibitem{Kolb:2003dz}
  P.~F.~Kolb and U.~W.~Heinz,
  in \textit{Quark-Gluon Plasma 3,} eds.\ R.~C.~Hwa and X.~N.~Wang
  (World Scientific, Singapore, 2004), p.\ 634
  [arXiv:nucl-th/0305084].

\bibitem{Gyulassy:2004zy}
  M.~Gyulassy and L.~McLerran,
  Nucl.\ Phys.\  A {\bf 750} (2005) 30
  [arXiv:nucl-th/0405013].

\bibitem{BraunMunzinger:2003zd}
  P.~Braun-Munzinger, K.~Redlich and J.~Stachel,
  in \textit{Quark-Gluon Plasma 3,} eds.\ R.~C.~Hwa and X.~N.~Wang
  (World Scientific, Singapore, 2004), p.\ 491
  [arXiv:nucl-th/0304013].

\bibitem{Adams:2005dq}
  J.~Adams {\it et al.}  [STAR Collaboration],
  Nucl.\ Phys.\  A {\bf 757} (2005) 102
  [arXiv:nucl-ex/0501009].

\bibitem{Adams:2003xp}
  J.~Adams {\it et al.}  [STAR Collaboration],
  Phys.\ Rev.\ Lett.\  {\bf 92} (2004) 112301
  [arXiv:nucl-ex/0310004].

\bibitem{Eskola:2002wx}
  K.~J.~Eskola, H.~Niemi, P.~V.~Ruuskanen and S.~S.~Rasanen,
  Phys.\ Lett.\  B {\bf 566} (2003) 187
  [arXiv:hep-ph/0206230].

\bibitem{Eskola:2005ue}
  K.~J.~Eskola, H.~Honkanen, H.~Niemi, P.~V.~Ruuskanen and S.~S.~Rasanen,
  Phys.\ Rev.\  C {\bf 72} (2005) 044904
  [arXiv:hep-ph/0506049].

\bibitem{Bebie:1991ij}
  H.~Bebie, P.~Gerber, J.~L.~Goity and H.~Leutwyler,
  Nucl.\ Phys.\ B {\bf 378,} (1992) 95.

\bibitem{Arbex:2001vx}
  N.~Arbex, F.~Grassi, Y.~Hama and O.~Socolowski,
  Phys.\ Rev.\ C {\bf 64,} (2001) 064906;
  W.~L.~Qian, R.~Andrade, F.~Grassi, O.~J.~Socolowski, T.~Kodama and Y.~Hama,
  Int.\ J.\ Mod.\ Phys.\  E {\bf 16} (2007) 1877
  [arXiv:nucl-th/0703078];
  W.~L.~Qian, R.~Andrade, F.~Grassi, Y.~Hama and T.~Kodama,
  arXiv:0709.0845 [nucl-th].

\bibitem{Teaney:2002aj}
  D.~Teaney,
  arXiv:nucl-th/0204023.

\bibitem{Hirano:2002ds}
  T.~Hirano and K.~Tsuda,
  Phys.\ Rev.\ C {\bf 66}, (2002) 054905
  [arXiv:nucl-th/0205043].

\bibitem{Kolb:2002ve}
  P.~F.~Kolb and R.~Rapp,
  Phys.\ Rev.\ C {\bf 67}, (2003) 044903
  [arXiv:hep-ph/0210222].

\bibitem{Adcox:2004mh}
  K.~Adcox {\it et al.}  [PHENIX Collaboration],
  Nucl.\ Phys.\  A {\bf 757} (2005) 184
  [arXiv:nucl-ex/0410003].

\bibitem{Hirano:2005wx}
  T.~Hirano and M.~Gyulassy,
  Nucl.\ Phys.\  A {\bf 769} (2006) 71
  [arXiv:nucl-th/0506049].

\bibitem{Dumitru:1998es}
  A.~Dumitru and D.~H.~Rischke,
  Phys.\ Rev.\  C {\bf 59}, (1999) 354
  [arXiv:nucl-th/9806003].

\bibitem{Eidelman:2004wy}
  S.~Eidelman {\it et al.}  [Particle Data Group],
  Phys.\ Lett.\  B {\bf 592} (2004) 1.

\bibitem{Huovinen:2005gy}
  P.~Huovinen,
  Nucl.\ Phys.\  A {\bf 761}, (2005) 296
  [arXiv:nucl-th/0505036].

\bibitem{Kolb:2000sd}
  P.~F.~Kolb, J.~Sollfrank and U.~W.~Heinz,
  Phys.\ Rev.\  C {\bf 62}, (2000) 054909
  [arXiv:hep-ph/0006129].

\bibitem{Berges:2005ai}
  J.~Berges, S.~Borsanyi and C.~Wetterich,
  Nucl.\ Phys.\  B {\bf 727}, (2005) 244
  [arXiv:hep-ph/0505182].

\bibitem{Kovchegov:2007pq}
  Y.~V.~Kovchegov and A.~Taliotis,
  Phys.\ Rev.\  C {\bf 76}, (2007) 014905
  [arXiv:0705.1234 [hep-ph]].

\bibitem{Mrowczynski:2005ki}
  S.~Mrowczynski,
  Acta Phys.\ Polon.\  B {\bf 37} (2006) 427
  [arXiv:hep-ph/0511052].

\bibitem{Kolb:2001qz}
  P.~F.~Kolb, U.~W.~Heinz, P.~Huovinen, K.~J.~Eskola and K.~Tuominen,
  Nucl.\ Phys.\  A {\bf 696}, (2001) 197
  [arXiv:hep-ph/0103234].

\bibitem{Muller:2005wi}
  B.~Muller,
  arXiv:nucl-th/0508062.

\bibitem{Heinz:2006ur}
  U.~Heinz and G.~Kestin,
  PoS C {\bf POD2006} (2006) 038
  [arXiv:nucl-th/0612105].

\bibitem{Karsch}
  F.~Karsch and E.~Laermann, 
  in \textit{Quark-Gluon Plasma 3,} eds.\ R.~C.~Hwa and X.~N.~Wang
  (World Scientific, Singapore, 2004), p.\ 1
  [arXiv:hep-lat/0305025].

\bibitem{Kaneta:2004zr}
  M.~Kaneta and N.~Xu,
  arXiv:nucl-th/0405068.

\bibitem{Andronic:2005yp}
  A.~Andronic, P.~Braun-Munzinger and J.~Stachel,
  Nucl.\ Phys.\  A {\bf 772}, (2006) 167
  [arXiv:nucl-th/0511071].

\bibitem{Baran:2003nm}
  A.~Baran, W.~Broniowski and W.~Florkowski,
  Acta Phys.\ Polon.\ B {\bf 35}, (2004) 779 
  [arXiv:nucl-th/0305075].

\bibitem{Cleymans:2004pp}
  J.~Cleymans, B.~Kampfer, M.~Kaneta, S.~Wheaton and N.~Xu,
  Phys.\ Rev.\ C {\bf 71}, (2005) 054901
  [arXiv:hep-ph/0409071].

\bibitem{Adler:2003cb}
  S.~S.~Adler {\it et al.}  [PHENIX Collaboration],
  Phys.\ Rev.\  C {\bf 69}, (2004) 034909
  [arXiv:nucl-ex/0307022].

\bibitem{Roy}
  Roy Lacey, private communication.

\bibitem{Back:2004zx}
  B.~B.~Back {\it et al.}  [PHOBOS Collaboration],
  Phys.\ Rev.\  C {\bf 70}, (2004) 051901
  [arXiv:nucl-ex/0401006].

\bibitem{Kataja:1990tp}
  M.~Kataja and P.~V.~Ruuskanen,
  Phys.\ Lett.\  B {\bf 243} (1990) 181.

\bibitem{Adams:2004bi}
  J.~Adams {\it et al.}  [STAR Collaboration],
  Phys.\ Rev.\  C {\bf 72}, (2005) 014904
  [arXiv:nucl-ex/0409033].

\bibitem{Phenix-v2}
  S.~S.~Adler {\it et al.}  [PHENIX Collaboration],
  Phys.\ Rev.\ Lett.\  {\bf 91}, (2003) 182301
  [arXiv:nucl-ex/0305013].

\bibitem{Chaudhuri:2007zm}
  A.~K.~Chaudhuri,
  arXiv:0704.0134 [nucl-th].

\bibitem{Romatschke:2007mq}
  P.~Romatschke and U.~Romatschke,
  Phys.\ Rev.\ Lett.\  {\bf 99} (2007) 172301
  [arXiv:0706.1522 [nucl-th]].

\bibitem{Song:2007fn}
  H.~Song and U.~W.~Heinz,
  Phys.\ Lett.\  B {\bf 658} (2008) 279
  [arXiv:0709.0742 [nucl-th]].

\bibitem{Hirano:2005xf}
  T.~Hirano, U.~W.~Heinz, D.~Kharzeev, R.~Lacey and Y.~Nara,
  Phys.\ Lett.\  B {\bf 636}, (2006) 299
  [arXiv:nucl-th/0511046].

\bibitem{Bass:2000ib}
  S.~A.~Bass and A.~Dumitru,
  Phys.\ Rev.\  C {\bf 61} (2000) 064909
  [arXiv:nucl-th/0001033].

\bibitem{Teaney:2001av}
  D.~Teaney, J.~Lauret and E.~V.~Shuryak,
  arXiv:nucl-th/0110037.

\bibitem{Nonaka:2006yn}
  C.~Nonaka and S.~A.~Bass,
  Phys.\ Rev.\  C {\bf 75}, (2007) 014902
  [arXiv:nucl-th/0607018].

\bibitem{Rapp:2000gy}
  R.~Rapp and E.~V.~Shuryak,
  Phys.\ Rev.\ Lett.\  {\bf 86} (2001) 2980
  [arXiv:hep-ph/0008326];
  R.~Rapp,
  Phys.\ Rev.\ C {\bf 66}, (2002) 017901
  [arXiv:hep-ph/0204131].

\bibitem{Cassing:2001ds}
  W.~Cassing,
  Nucl.\ Phys.\  A {\bf 700} (2002) 618
  [arXiv:nucl-th/0105069].

\bibitem{NoronhaHostler:2007fg}
  J.~Noronha-Hostler, C.~Greiner and I.~A.~Shovkovy,
  arXiv:nucl-th/0703079.

\bibitem{Pratt:1999ku}
  S.~Pratt and K.~Haglin,
  Phys.\ Rev.\ C {\bf 59}, (1999) 3304.

\bibitem{Huovinen:2003sa}
  P.~Huovinen and J.~I.~Kapusta,
  Phys.\ Rev.\ C {\bf 69}, (2004) 014902
  [arXiv:nucl-th/0310051].

\end{thebibliography}
\end{document}